\documentclass[11pt]{article}
\newcommand{\dd}{\mathrm{d}}
\newcommand{\ii}{\mathrm{i}}

\usepackage{graphicx}
\usepackage{amsmath}
\usepackage{amssymb}
\usepackage{amsxtra}
\usepackage{cite}

\begin{document}

\title{Gravitational Interactions and\\ Fine-Structure Constant}
\author{U. D. Jentschura and J. H. Noble\\
{\it Department of Physics, Missouri University of}\\
{\it Science and Technology, Rolla, Missouri 65409, USA}\\
{\bf \small ulj@mst.edu}\\[2ex]
I. N\'andori\\
{\it MTA--DE Particle Physics Research Group,}\\
{\it P.O.Box 51, H--4001 Debrecen, Hungary}}

\maketitle

\begin{abstract}
Electromagnetic and gravitational central-field problems
are studied with relativistic quantum mechanics on curved space-time
backgrounds. Corrections to the transition current are identified.
Analogies of the gravitational and electromagnetic spectra
suggest the definition of a gravitational fine-structure constant.
The electromagnetic and gravitational coupling constants 
enter the Einstein--Hilbert--Maxwell Lagrangian. 
We postulate that  the variational principle holds with regard to a global 
dilation transformation of the space-time coordinates.
The variation suggests is consistent with a  functional relationship
of the form $\alpha_{\rm QED} \propto (\alpha_G)^{1/2}$,
where $\alpha_{\rm QED}$ is the electrodynamic fine-structure
constant, and $\alpha_G$ its gravitational analogue.
\end{abstract}

%
% Introduction
%
\section{Introduction}
\label{sec1}

If we are ever to gain a better understanding of the relationship
of gravitational interactions and electrodynamics 
in the quantum world, then a very practical 
approach is to try to solve a number of important example 
problems in gravitational theory, whose solution is 
known in electromagnetic theory, 
to try to generalize the approach to the gravitational 
analogue, and to compare.
In order to proceed, it is not necessarily required to 
quantize space-time itself~\cite{Je2014dirac}. Indeed, the 
formulation of quantum mechanics on curved-space 
backgrounds in itself constitutes an interesting 
problem~\cite{BrWh1957,Bo1975prd,SoMuGr1977,ZaMB2012,GiLi2014}.
A priori, one might think that the simple
substitution $\partial/\partial x^i \to 
\nabla_i$ is the Schr\"{o}dinger equation 
might suffice. Here, $\partial/\partial x^i $ is 
the $i$th partial derivative with respect to the 
$i$th spatial coordinate, whereas $\nabla_i$  is the $i$th 
covariant derivative. 
However, this naive approach is destined to fail;
the gravitational theory of Einstein and Hilbert inherently 
is a relativistic theory, and the only way to 
describe quantum particles on curved space-times is 
to start from a fully relativistic wave function.
The Dirac equation
\begin{equation}
\left( \ii \gamma^\mu \partial_\mu - m \right) \psi(x) = 0
\end{equation}
generalizes as follows to a curved space-time 
background~\cite{BrWh1957,Bo1975prd,SoMuGr1977,ZaMB2012,GiLi2014},
\begin{equation}
\left( \ii \gamma^\mu(x) (\partial_\mu -\Gamma_\mu) - m \right) \psi(x) = 0 \,.
\end{equation}
The Dirac algebra~\cite{Di1928a,Di1928b,JeWu2012jpa} needs to be generalized to 
the local metric $g^{\mu\nu}(x)$,
\begin{equation}
\{ \gamma^\mu(x), \gamma^\nu(x) \} = 2 \, g^{\mu\nu}(x)  \,,
\qquad
\sigma^{\mu\nu}(x) = \frac{\ii}{2} [\gamma^\mu(x), \gamma^\nu(x)] \,.
\end{equation}
The spin connection matrix $\Gamma_\mu$ is given as 
\begin{equation}
\Gamma_\mu = - \frac{\ii}{4} \, g_{\rho\alpha}(x) \,
\left( \frac{\partial {b_\nu}^\beta(x)}{\partial x^\mu} \,
{a^\alpha}_\beta(x) - {\Gamma^\alpha}_{\nu \mu} \right) \,
\sigma^{\rho\nu}(x) \,,
\end{equation}
where repeated indices are summed.
Finally, the $a$ and $b$ coefficients belong to the
square root of the metric,
\begin{equation}
\gamma_\rho(x) = {b_\rho}^\alpha(x) \, \gamma_\alpha\,,
\qquad
\gamma^\alpha(x) = {a^\alpha}_\rho(x) \, \gamma^\rho,
\end{equation}
where the $\gamma^\alpha$ are the flat-space Dirac matrices,
which are preferentially used in the Dirac 
representation~\cite{JeWu2012jpa,JeNo2013pra,JeNo2014jpa,Je2014dirac,Je2014pra}.
The Christoffel symbols are
${\Gamma^\alpha}_{\nu \rho} \equiv {\Gamma^\alpha}_{\nu \rho}(x)$.

%
% Central--Field Problem
%
\section{Central--Field Problem}
\label{sec2}

%
% Foldy--Wouthuysen Method
%
\subsection{Foldy--Wouthuysen Method}

The Foldy--Wouthuysen method~\cite{FoWu1950,BjDr1964} is a standard tool for
the extraction of the physical, nonrelativistic degrees of freedom, from a
fully relativistic Dirac theory.  The general paradigm is as follows: The
positive and negative energy solutions of a (generalized) Dirac equation are
intertwined in the fully relativistic formalism.  One has to separate the upper
and lower spinors in the bispinor solution, and in order to do so, one
eliminates the ``off-diagonal couplings'' of the upper and lower spinor
components order by order in some perturbative parameters, possibly, using
iterated (unitary) transformations.

For the plain free Dirac Hamiltonian, a standard method 
exists to all orders in perturbation theory,
while for more difficult problems, one manifestly 
has to resort to a perturbative formalism~\cite{FoWu1950,BjDr1964}. 
A suitable expansion parameter in a general
case is the particle's momentum operator.
Let us consider a space-time metric of the form
\begin{equation}
\label{vw}
{\overline g}_{\mu\nu} =
{\rm diag}\left( w^2(r), -v^2(r), -v^2(r), -v^2(r) \right) \,.
\end{equation}
The Schwarzschild metric in
isotropic coordinates (see Sec.~43 of Chap.~3 of Ref.~\cite{Ed1924}),
involves the Schwarzschild radius $r_s$,
\begin{align}
w =&\;
\left( 1 - \frac{r_s}{4 r} \right) \, \left( 1 + \frac{r_s}{4 r} \right)^{-1} =
\frac{4 r - r_s}{4 r + r_s} \approx 1 - \frac{r_s}{2 r} \,,
\nonumber\\[1.0ex]
v =& \; \left( 1 + \frac{r_s}{4 r} \right)^2 \approx 1 + \frac{r_s}{2 r} \,,
\qquad
\frac{w}{v} = \frac{16\,r^2\,(4 r - r_s)}{(4 r + r_s)^3} \approx
1 - \frac{r_s}{r} \,.
\end{align}
The Schwarzschild radius reads as
$r_s = 2 \, G \, M$, where $G$ is Newton's gravitational constant,
and $M$ is the mass of the planet (or ``black hole'').
The Hamiltonian or time translation operator 
is necessarily ``noncovariant'' in the sense that the 
time coordinate needs to be singled out.
If we insist on using the time translation with respect
to the time coordinate $\dd t$ in the metric 
$\dd s^2 = w^2(r) \, \dd t^2 - v^2(r) \dd \vec r^2$ and 
bring the Hamiltonian into Hermitian form
[see Ref.~\cite{Ob2001} and Eqs.~(9)--(13) of Ref.~\cite{JeNo2013pra}],
then we obtain
\begin{equation}
\label{HDS}
H_{\rm DS} =
\frac12 \, \left\{ \vec\alpha \cdot \vec p,
\left( 1 - \frac{r_s}{r} \right) \right\} +
\beta m \left( 1 - \frac{r_s}{2 r} \right) \,,
\end{equation}
where $\alpha^i = \gamma^0 \, \gamma^i$ is the 
Dirac $\alpha$ matrix (we here use the Dirac representation).
The Foldy--Wouthuysen transformed Dirac--Schwarzschild Hamiltonian
is finally obtained as~\cite{JeNo2013pra}
\begin{align}
\label{HFW}
& H_{\rm FW} =
 \beta \, \left( m + \frac{\vec p^{\,2}}{2 m} -
\frac{\vec p^{\,4}}{8 m^3} \right)
- \beta \, \frac{m \, r_s}{2 \, r}
\\[1.0ex]
& \; + \beta \, \left(
- \frac{3 r_s}{8 m} \,
\left\{ \vec p^{\,2}, \frac{1}{r} \right\}
+ \frac{3 \pi r_s}{4 m} \delta^{(3)}(\vec r) \,
+ \frac{3 r_s}{8 m} \, \frac{\vec\Sigma \cdot \vec L}{r^3} \right) \,.
\nonumber
\end{align}
The parity-violating terms obtained in 
Refs.~\cite{DoHo1986,Ob2001} are spurious.

\begin{figure}[t!]
\begin{center}
\includegraphics[width=0.45\linewidth]{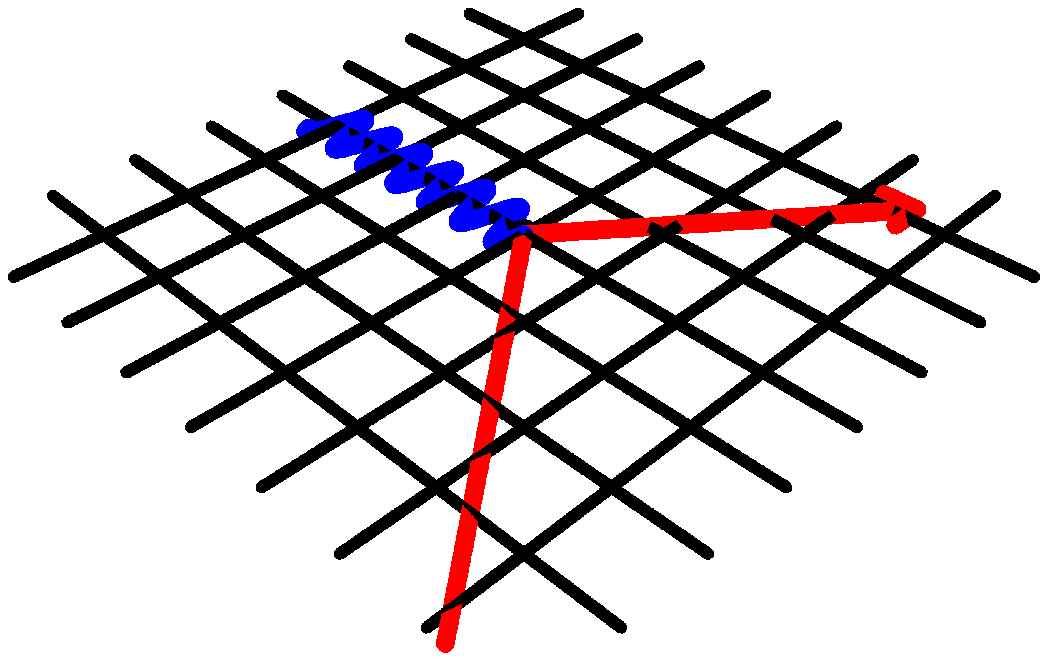}
\hfill
\includegraphics[width=0.45\linewidth]{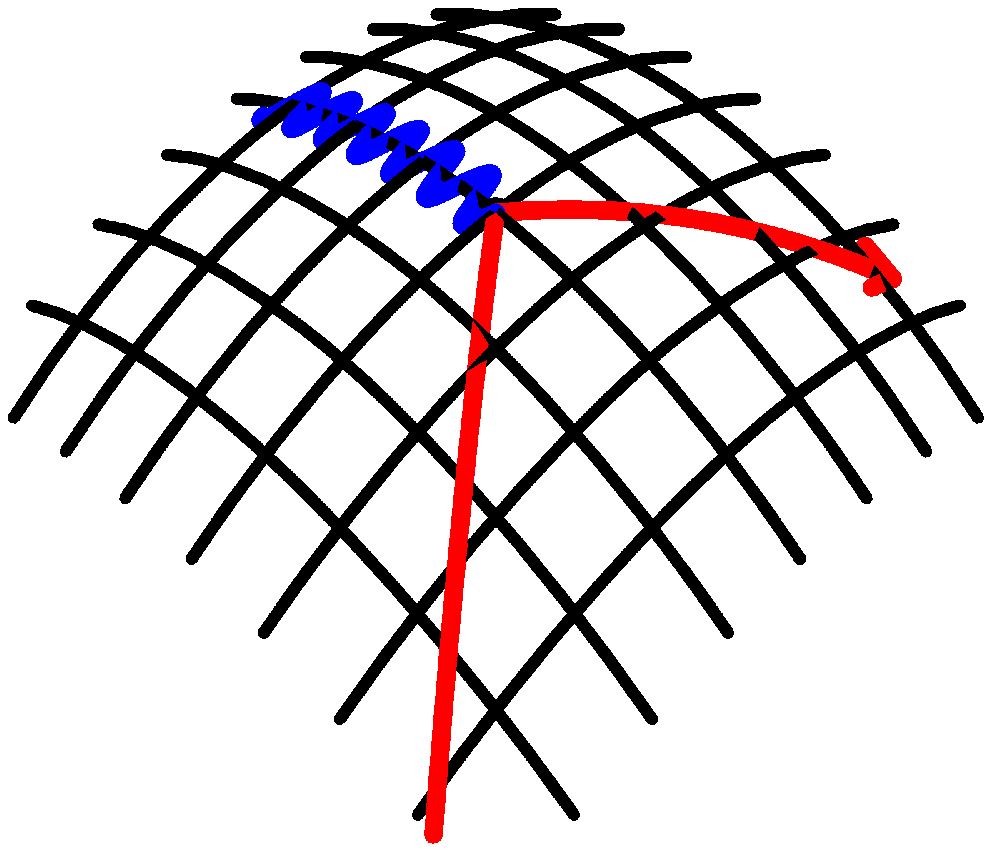}
\end{center}
\caption{\label{fig1}The flat-space photon emission vertex 
(left figure) is promoted 
to a curved-space vertex (right figure) 
in general relativity. The curved background 
leads to higher-order corrections to the transition current,
which are summarized, for the Schwarzschild metric,
in Eq.~\eqref{jFW}.}
\label{fig1label}
\end{figure}

%
% Transition Current
%
\subsection{Transition Current}

As we couple the Dirac--Schwarzschild Hamiltonian~\eqref{HDS} 
to an electromagnetic field (see Fig.~\ref{fig1}), 
it is clear that the transition current 
in the interaction Hamiltonian is $H_{\rm int} = -\vec j \cdot \vec A$.
takes the form
\begin{equation}
\label{ji}
j^i = \frac12 \, \left\{ 1 - \frac{r_s}{r}, 
\alpha^i \, \exp(\ii \vec k \cdot r) \right\}  \,.
\end{equation}
We now employ the multipole expansion 
\begin{equation}
\alpha^i \, \exp(\ii \vec k \cdot r) \approx
\alpha^i + \alpha^i \, (\ii \vec k \cdot \vec r) -
\frac12 \alpha^i (\vec k \cdot r)^2
\end{equation}
A unitary transformation with the 
same generators are used for the Dirac--Schwarzschild 
Hamiltonian then yields the result~\cite{JeNo2013pra},
\begin{align}
\label{jFW}
j_{\rm FW}^i = & \;
\frac{p^i}{m} - \frac{p^i \, \vec p^{\,2}}{2 m}
- \frac{\ii}{2 m} \left( \vec k \times \vec\sigma \right)^i
+ \frac12 \, \left\{ \frac{p^i}{m}, \, (\ii \vec k \cdot \vec r) \right\}
\nonumber\\[1.0ex]
& \; 
-\frac{1}{4} \, \left\{ (\vec k \cdot \vec r)^2, \frac{p^i}{m} \right\}
+ \frac{1}{2 m} \left( \vec k \cdot \vec r \right) \, (\vec k \times \vec\sigma)^i 
\nonumber\\[1.0ex]
& \; 
- \frac{3}{4} \, \left\{ \frac{p^i}{m}, \frac{r_s}{r} \right\} 
+ \frac{r_s}{2 r} \frac{(\vec\sigma \times \vec r)^i}{m \, r^2} 
- \frac12 \, \left\{ \left( \ii \vec k \cdot \vec r \right), 
\left\{ \frac{p^i}{m}, \frac{r_s}{r} \right\} \right\} 
\nonumber\\[1.0ex]
& \; 
+ \frac{3 \ii r_s}{4 r} \frac{( \vec k \times \vec \sigma )^i}{m} 
+ \frac{1}{4} \, \left\{ \frac{r_s}{r} \, (\ii \vec k \cdot \vec r), 
\frac{p^i}{m} \right\} \,.
\end{align}
This result contains a gravitational kinetic correction,
and gravitational corrections to the 
magnetic coupling, in addition to the known
multipole and retardation corrections~\cite{BjDr1964,JePa1996}.

%
% Spectrum
%
\subsection{Spectrum}

The bound-state spectrum resulting from the Hamiltonian~\eqref{HDS}
has recently been evaluated as~\cite{Je2014pra},
\begin{align}
\label{spectrum}
E_{n \ell j} =& \; -\frac{\alpha_G^2 m_e c^2}{2 n^2} +
\alpha^4_G m_e c^2 \,
\left( \frac{15}{8 n^4} \right.
\\[0.133ex]
& \; \left. - \frac{(7 j + 5)\, \delta_{\ell, j+1/2}}{(j+1) \, (2j+1) \, n^3}
- \frac{(7 j + 2)\, \delta_{\ell, j-1/2}}{j \, (2j+1) \, n^3} \right)
\nonumber\\[0.133ex]
=& -\frac{\alpha_G^2 m_e c^2}{2 n^2} +
\frac{\alpha^4_G m_e c^2}{n^3} \, \left( \frac{15}{8 n} -
\frac{14 \, \varkappa + 3}{2 \,|\varkappa| \, (2 \varkappa + 1)} \right)\,,
\nonumber
\end{align}
where $\ell$ is the orbital angular momentum, 
$j$ is the total angular momentum of the bound particle,
and $\varkappa$ is the (integer) Dirac angular quantum number,
\begin{equation}
\label{varkappa}
\varkappa = 2 (\ell - j) \, (j + 1/2)
= (-1)^{j+\ell +1/2} \, \left(j + \frac12 \right) \,.
\end{equation}
For a bound electron-proton system, the 
coupling constant entering the gravitational spectrum 
given in Eq.~\eqref{spectrum} reads as 
\begin{equation}
\label{alphaG}
\alpha_G = \frac{G \, m_e \, m_p}{\hbar \, c} =
3.21637(39) \times 10^{-42} \,.
\end{equation}
The coupling $\alpha_G $ is much larger than 
for particles bound to macroscopic objects.
By contrast, the electrodynamic coupling parameter
\begin{equation}
\alpha_{\rm QED} = \frac{e^2}{4 \pi \hbar \epsilon_0 c} 
\approx \frac{1}{137.036} 
\end{equation}
is just the fine-structure constant.

%
% Global Dilation Transformation
%
\section{Global Dilation Transformation}
\label{sec3}

%
% Lagrangian
%
\subsection{Lagrangian}

The analogy of the leading (Schr\"{o}dinger) term in 
Eq.~\eqref{spectrum} for the nonrelativistic contribution
to the bound-state energy
(under the replacement $\alpha_G \to \alpha_{\rm QED}$)
may encourage us to look for connections of 
gravitational and electromagnetic interactions 
on a more global scale, possibly, using scaling 
transformations~\cite{JeNa2014epjh}.
Indeed, the first attempts to unify electromagnetism with gravity 
are almost 100~years old~\cite{Ka1921,Kl1926}.
Let us apply a scaling transformation to the 
boson and fermion fields,
\begin{equation}
A^\mu \to \lambda \, A^\mu \,,
\qquad
A_\mu \to \lambda \, A_\mu \,,
\qquad
\psi \to \lambda \, \psi \,,
\end{equation}
combined with a transformation of the coordinates,
\begin{equation}
x^\mu \to \lambda^{-{1/2}} \, x^\mu \,,
\qquad
x_\mu \to \lambda^{-1/2} \, x_\mu \,,
\end{equation}
and of the metric
\begin{equation}
g_{\mu\nu} \to \lambda g_{\mu\nu} \,,
\qquad
g^{\mu\nu} \to \lambda^{-1} \, g^{\mu\nu} \,,
\end{equation}
Under this transformation, 
the space-time intervals,
the integration measure,
the Ricci tensor $R_{\mu\nu}$
and the curvature scalar $R$, 
transform as follows,
\begin{subequations}
\begin{align}
\dd s^2 =& \; g_{\mu\nu} \, \dd x^\mu \, \dd x^\nu = 
g^{\mu\nu} \, \dd x_\mu \, \dd x_\nu 
\to \dd s^2 \,,
\\[0.133ex]
\dd^4 x =& \; \dd^4 x \to \frac{\dd^4 x}{\lambda^2} \,,
\qquad
\qquad
\det \, g = \det \, g_{\mu\nu} \to \lambda^4 \, \det \, g \,,
\\[0.133ex]
R_{\mu\nu} \to & \; \lambda \, R_{\mu\nu}  \,,
\qquad
\qquad
R = g^{\mu\nu} \, R_{\mu\nu} \to R \,.
\end{align}
\end{subequations}
The Einstein--Maxwell Lagrangian, 
with a coupling to the fermion terms, is given as
\begin{align}
\label{Sorig}
S = & \; \int \dd^4 x \, \sqrt{- {\rm det} \, {g}} \,
\left\{ \frac{R}{16 \, \pi \, G} - \frac14 \, F^{\mu\nu} \, F_{\mu\nu} 
\right.
\nonumber\\[0.133ex]
& \; \left. + {\overline \psi}(x) \; \left[ \ii {\overline \gamma}^\mu \;
\left( \nabla_\mu - e \, A_\mu \right) - m \right] \, \psi(x) \right\}  \,.
\end{align}
It transforms into
\begin{align}
S' =& \; \int \frac{\dd^4 x}{\lambda^2} \, 
\sqrt{- \lambda^4 \, {\rm det} \, {g}} \,
\left\{ \frac{R}{16 \, \pi \, G} - \frac{ \lambda^2}{4} \, F^{\mu\nu} \, F_{\mu\nu} 
\right.
\nonumber\\[0.133ex]
& \; \left.
+ \lambda^{2} \; {\overline \psi}(x) \; 
\left[ \ii \, \lambda^{-1/2} \, 
{\overline \gamma}^\mu \;
\left( \lambda^{1/2} \nabla_\mu - e \, \lambda \, A_\mu \right) 
- m \right] \, \psi(x) \right\} \,,
\end{align}
which can be rearranged into
\begin{align}
\label{S2prime2}
S'' =& \frac{S'}{\lambda^2} = \int \dd^4 x \, \sqrt{- {\rm det} \, {g}} \,
\left\{ \frac{R}{16 \, \pi \, G \, \lambda^2} - 
\frac{1}{4} \, F^{\mu\nu} \, F_{\mu\nu} 
\right.
\nonumber\\[0.133ex]
& \; \left. 
+ {\overline \psi}(x) \; \left[ \ii {\overline \gamma}^\mu \;
\left( \nabla_\mu - e \, \lambda^{1/2} \, A_\mu \right) - m \right] \, \psi(x)
\right\} \,.
\end{align}
The Lagrangian $S''$ is the same $S$, but with scaled coupling constants,
\begin{equation}
\label{scaling}
G \to \lambda^2 \, G \,,
\qquad
e^2 \to \lambda \, e^2 \,.
\end{equation}
This scaling suggests a deeper connection of the 
coupling constants of electromagnetic and gravitational
interactions, which is explored in further detail in Ref.~\cite{JeNa2014epjh}.

%
% Coupling Constants
%
\subsection{Coupling Constants}

If we assume that the scaling~\eqref{scaling} holds globally, 
with the current Universe ``picking'' a value of $\lambda$,
then this scaling might suggest a relationship of the type
\begin{equation}
\alpha_{\rm QED}^2 \propto 
e^4 \propto \lambda^2 \propto G \,.
\end{equation}
Indeed, as discussed in Ref.~\cite{JeNa2014epjh},
a relationship of the type $\alpha_{\rm QED} \propto \sqrt{G}$
is otherwise suggested by string theory;
the rough analogy being that gravitational interactions 
in string theory correspond to ``closed'' strings 
while electromagnetic interactions correspond to ``open'' strings.
The product of two ``open'' string amplitudes is 
proportional to $e^2 \propto g_0^2 \propto \alpha_{\rm QED}$, 
while the ``closed''-string amplitude is 
proportional to $\kappa \propto g_c \propto \sqrt{G}$.
According to Eq.~(3.7.17) of Ref.~\cite{Po1998vol1}, the 
proportionality 
\begin{equation}
g_o^2 \propto g_c \qquad
\Leftrightarrow \qquad
\alpha_{\rm QED}^2 \propto \sqrt{G}
\end{equation}
therefore is suggested by string theory.
A simple analytic form of the proportionality factor in the 
relationship $\alpha_{\rm QED}^2 \propto \sqrt{G}$
has recently been given in Eq.~(8) of Ref.~\cite{Je2014dirac}.

%
% Conclusions
%
\section{Conclusions}
\label{sec4}

We have performed an analysis of the gravitationally
coupled Dirac equation in the curved space-time 
surrounding a central gravitating object,
which is described by the (static) Schwarzschild metric.
The Foldy--Wout\-huysen method leads to 
gravitational zitterbewegung terms and the 
gravitational spin-orbit coupling, which is also 
known as the Fokker precession term.
In a curved space-time, the photon emission 
vertex receives additional corrections due to 
the curved background, which can be given, within the 
multipole expansion and for a conceptually simple
background metric (e.g., the Schwarzschild metric),
in closed analytic form (at least for the first 
terms of the multipole and retardation expansion).
The gravitational bound states display a certain
analogy for the gravitational as compared to the 
electromagnetic (Schr\"{o}dinger) central-field problem.
Based on this analogy, one may explore 
possible connections of the gravitational and 
electromagnetic coupling constants, based on scaling 
arguments. Such a scaling transformation
gives additional support for the relationship
$\alpha_{\rm QED}^2 \propto \sqrt{G}$,
which has been suggested by string theory~\cite{Po1998vol1}.

\section*{Acknowledgements}

The research has been supported by the National Science Foundation
(Grants PHY--1068547 and PHY--1403973).

\end{document}